\documentclass[journal=jacsat,manuscript=article]{achemso}

\usepackage[version=3]{mhchem} 
\usepackage[usenames,dvipsnames]{xcolor}
\usepackage{float} 
\usepackage{gensymb}

\author{Kamal Choudhary}
 \email{kchoudh2@jhu.edu}
 \affiliation{%
 Material Measurement Laboratory, National Institute of Standards and Technology,
Gaithersburg, MD 20899, USA 
}
\altaffiliation{%
Department of Materials Science and Engineering, Whiting School of Engineering, The Johns Hopkins University, Baltimore, MD 21218, USA
}%
\alsoaffiliation{%
Department of Electrical and Computer Engineering, Whiting School of Engineering, The Johns Hopkins University, Baltimore, MD 21218, USA
}%

\title{DiffractGPT: Atomic Structure Determination from X-ray Diffraction Patterns using Generative Pre-trained Transformer}

\abbreviations{IR,NMR,UV}

\begin{document}

\begin{tocentry}

\begin{center}
\includegraphics[width=6cm]{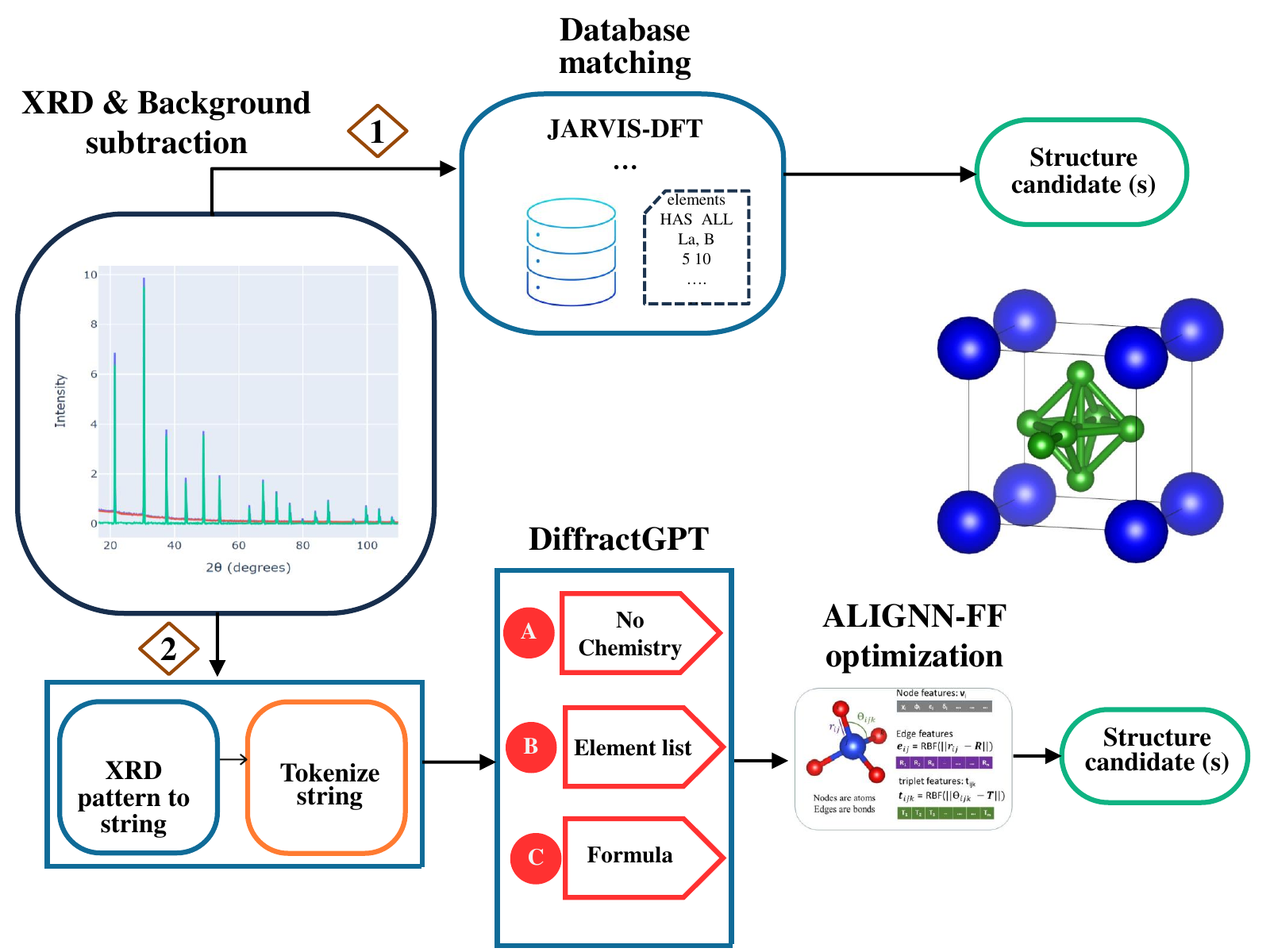}
\label{TOC}
\end{center}
\end{tocentry}

\begin{abstract}

Crystal structure determination from powder diffraction patterns is a complex challenge in materials science, often requiring extensive expertise and computational resources. This study introduces DiffractGPT, a generative pre-trained transformer model designed to predict atomic structures directly from X-ray diffraction (XRD) patterns. By capturing the intricate relationships between diffraction patterns and crystal structures, DiffractGPT enables fast and accurate inverse design. Trained on thousands of atomic structures and their simulated XRD patterns from the JARVIS-DFT dataset, we evaluate the model across three scenarios: (1) without chemical information, (2) with a list of elements, and (3) with an explicit chemical formula. The results demonstrate that incorporating chemical information significantly enhances prediction accuracy. Additionally, the training process is straightforward and fast, bridging gaps between computational, data science, and experimental communities. This work represents a significant advancement in automating crystal structure determination, offering a robust tool for data-driven materials discovery and design.

\end{abstract}













Since the discovery of X-rays in 1895, they have been widely used in medical imaging, crystallography, and astronomy \cite{als2011elements}. Numerous experimental techniques in materials science rely on X-rays, including X-ray diffraction (XRD), X-ray fluorescence (XRF), X-ray photoelectron spectroscopy (XPS), small-angle X-ray scattering (SAXS), X-ray tomography (XRT), X-ray reflectometry (XRR), grazing incidence X-ray diffraction (GIXRD), and resonant inelastic X-ray scattering (RIXS) \cite{giacovazzo2002fundamentals,wyon2010x}. Among these, XRD plays a crucial role in determining atomic structures and uncovering the mechanisms underlying mechanical strength, electronic properties, optical behavior, and chemical reactivity \cite{holder2019tutorial,brown2012x}. However, crystal structure determination currently involves extensive trial and error as well as expert knowledge. The main challenge lies in the reduction of chemical and three-dimensional structural information into one-dimensional diffraction patterns, which causes the loss of phase information and complicates structure determination. 

Additionally, the presence of peaks in the diffraction data of newly discovered compounds, complex materials, or multi-phase systems further exacerbates this challenge. Over the past few decades, Rietveld refinement, simulated annealing, and evolutionary algorithms have been developed to address this problem by iteratively fitting data to potential candidate structures \cite{giacovazzo2002fundamentals,wyon2010x}. Several widely used software tools, such as FullProf \cite{rodriguez1998fullprof}, the General Structure Analysis System (GSAS) \cite{larson1985gsas}, GenX \cite{glavic2022genx}, TOtal Pattern Analysis Solutions (TOPAS) \cite{coelho2018topas}, and Materials Analysis Using Diffraction (MAUD) \cite{wenk2010rietveld}, are available for this purpose. While these methods have been successful, they often require significant domain expertise, computational resources, and manual intervention, particularly when dealing with ambiguous or incomplete data.

In recent years, machine learning has emerged as a powerful tool in materials science, offering the potential to accelerate materials discovery and characterization \cite{choudhary2022recent,vasudevan2019materials,schmidt2019recent}. In particular, high-throughput materials design and process modeling, which are key driving forces behind the Materials Genome Initiative and the Creating Helpful Incentives to Produce Semiconductors (CHIPS) Act \cite{nistCHIPSGov}, require a bridge between experiments and multi-scale modeling components, where large language models (LLMs) could play a significant role. Moreover, two recent Nobel Prizes in Physics and Chemistry in 2024 for neural networks and AlphaFold clearly demonstrate the wide applicability of AI/ML in scientific research.

The AI/ML techniques have been successfully used for both forward (structure to property) and inverse (property to structure) tasks in materials design \cite{choudhary2022recent}. Generating crystal structures from XRD can be considered a generative AI-based inverse design task. Recent advancements in machine learning related to X-ray diffraction \cite{surdu2023x} include works by Park et al. \cite{park2017classification}, NeuralXRD \cite{zhdanov2023machine}, XRD\_is\_All\_You\_Need \cite{lee2022powder}, Crystallography Companion Agent (XCA) \cite{banko2021deep}, ARiXD-ML \cite{yanxon2023artifact}, Zaloga et al. \cite{zaloga2020crystal}, XTEC \cite{venderley2022harnessing}, Li et al. \cite{lee2020deep}, Maffettone et al. \cite{maffettone2021crystallography}, Oviedo et al. \cite{oviedo2019fast} and several others \cite{chen2024crystal,xin2023machine,xin2023advancements}. These works demonstrate the application of ML models for a wide range of tasks, including crystal lattice and space group classification, peak detection, and structure generation. In particular, the application of deep generative models such as Variational Autoencoders (VAEs) and Generative Adversarial Networks (GANs) has demonstrated the ability to generate complex atomic structures based on insights.

The potential of GPT in natural language processing (NLP), such as ChatGPT, has spurred interest in their applications beyond textual data, particularly in domains such as chemistry and materials science. The success of AtomGPT (Atomistic Generative Pre-trained Transformer) \cite{choudhary2024atomgpt} , which demonstrated the capability to generate atomic structures and predict material properties using transformer-based architectures, highlights the power of transformer models in handling materials data. AtomGPT establishes the relationship between atomic configurations as text and material properties, allowing it to tackle both forward and inverse design problems.

The GPT is a type of LLM originally developed for natural language processing and has demonstrated remarkable success in generating coherent and contextually relevant text \cite{vaswani2017attention,tunstall2022natural,rothman2021transformers}. Models such as ChatGPT \cite{wu2023brief} have been used for code generation, debugging, literature reviews, and numerous other tasks. However, if we attempt to perform forward/inverse materials design tasks, the outcomes can be quite poor \cite{pimentel2023challenging,jablonka2024leveraging,polak2024extracting}. Nevertheless, inspired by its simplicity of use and the massive success of ChatGPT, an alternate model, AtomGPT, was introduced, tailored for forward and inverse materials design.

While AtomGPT enables scalar material properties to be generated from atomic structures, its application for generating atomic structures from experimental properties, such as XRD, has not yet been explored. Based on these developments, we introduce \textcolor{black}{DiffractGPT (DGPT) }, a specialized generative model designed to directly predict crystal structures from powder X-ray diffraction (PXRD) patterns. DiffractGPT leverages the powerful architecture of AtomGPT, adapting it to the unique challenges of PXRD-based crystal structure determination. By training on large datasets such as JARVIS-DFT (JDFT), which comprises simulated PXRD patterns alongside their corresponding atomic structures, DiffractGPT learns to map complex diffraction data to accurate crystal structures. This approach enables the direct prediction of atomic arrangements from diffraction data, significantly reducing the need for iterative fitting and manual intervention. We further evaluate various application scenarios for DiffractGPT, such as XRD with no known chemical constituents, with guessed elements, and with explicit chemical formulas for structure design tasks. We also provide a web framework and tools to match the XRD patterns with existing data, as well as to generate new structures using the generative models. Most importantly, although we apply the models to XRD data, they can also be useful for other experiments, such as neutron and electron diffraction and other spectroscopic experiments.

The Joint Automated Repository for Various Integrated Simulations (JARVIS) - density functional theory (DFT) \cite{wines2023recent,choudhary2020joint} database used in this work contains nearly 80,000 bulk 3D materials and 1,100 2D materials. The JARVIS-DFT project originated about six years ago and has amassed millions of material properties, along with carefully converged atomic structures using tight convergence parameters and various exchange-correlation functionals. JARVIS-DFT encompasses a wide range of material classes, including metallic, semiconducting, insulating, superconducting, high-strength, topological, solar, thermoelectric, piezoelectric, dielectric, two-dimensional, magnetic, porous, defect, and various other types of bulk materials.

In this paper, we describe the architecture and training methodology of DiffractGPT and evaluate its performance on the PXRD dataset. DiffractGPT uses transformer architecture based on the Mistral AI model  \cite{jiang2023mistral} but can be easily adapted to other LLMs as well. We demonstrate that DiffractGPT not only matches the accuracy of traditional methods but also significantly reduces the computational time and expertise required for crystal structure determination. AtomGPT and DiffractGPT are analogous to AlphaFold (mentioned above) in their approach to solving complex structure-property relationships using machine learning. They adapt generative predictive frameworks to tackle fundamental challenges in materials science, mirroring what AlphaFold \cite{jumper2021highly} has achieved for biology. The results show the promise of using generative machine learning models for automating the crystal structure determination process, opening up new avenues for materials discovery and design. The code used in this study will be made available on the AtomGPT GitHub page: \url{https://github.com/usnistgov/atomgpt}.


The dataset used for this work is taken from the JARVIS-DFT database, which includes nearly 80,000 atomic structures and several material properties derived from density functional theory and powder X-ray diffraction patterns \cite{choudhary2020joint,wines2023recent,choudhary2018computational}. From an atomic structure and a given X-ray wavelength (here Cu K$\alpha$), the corresponding PXRD patterns can be easily calculated. The PXRD pattern was computed from the atomic structure by first calculating the reciprocal lattice vectors and interplanar spacings \(d_{hkl}\) for each set of Miller indices \((hkl)\). Bragg's law, \(n\lambda = 2d_{hkl} \sin \theta\), was used to convert these \(d\)-spacings into scattering angles \(2\theta\). The structure factor \(F(hkl)\) for each reflection was then calculated as the sum of atomic scattering contributions from all atoms in the unit cell, taking into account their positions and associated phase shifts. The atomic scattering factor \(f(\theta)\), which varies with the scattering angle, was used to model the electron density distribution around each atom accurately. The diffraction intensity for each reflection was obtained using the relation \(I(hkl) \propto |F(hkl)|^2\). A Gaussian broadening function was also applied to account for experimental resolution effects. The final XRD pattern was generated by summing the corrected intensities over all relevant reflections. All calculations were performed using custom scripts in the JARVIS-tools package to simulate the diffraction patterns for comparison with experimental data.

Such XRD predictions were carried out for all the data in the \textcolor{black}{JARVIS-DFT (JDFT)} dataset. The XRD dataset was split into a 90:10 ratio for training and testing the DiffractGPT models. This requires fine-tuning LLM models such as Mistral AI  \cite{jiang2023mistral}, which are based on transformer architecture. Each transformer block contains two main components: a multi-head self-attention mechanism and a position-wise feed-forward network. The input to the model is a sequence of tokens, which are first converted into embeddings and then passed through the transformer blocks. The scaled dot-product attention used in a transformer model can be written as:

\begin{equation}
   \text{Attention}(Q, K, V) = \text{softmax}\left(\frac{QK^T}{\sqrt{d_k}}\right)V
\end{equation}
where \(Q\), \(K\), and \(V\) represent the query, key, and value matrices, respectively. Here, \(d_k\) is the dimensionality of the key vectors. The multi-head attention is obtained by concatenating multiple such attention heads. The multi-head self-attention mechanism allows the model to focus on different parts of the input sequence when computing the output for a particular token.

There are thousands of LLMs, especially transformer models, that are publicly available. In particular, we use the Mistral AI 7 billion parameter model \cite{jiang2023mistral}, which employs Low-Rank Adaptation (LoRA) for parameter-efficient fine-tuning (PEFT) \cite{hu2021lora} adopted from the UnslothAI package \cite{unsloth}. Mistral is a powerful model with 7.3 billion parameters and has been shown to outperform the Large Language Model Meta AI (LLaMA) 2 13B \cite{touvron2023llama2}, LLaMA 1 34B \cite{touvron2023llama}, and ChatGPT \cite{wu2023brief} on several publicly available benchmarks. The Mistral 7B model combines efficiency and performance within a 7 billion parameter architecture. It introduces several key innovations, including Grouped-Query Attention for reduced computational complexity, Sliding Window Attention for processing longer sequences, and Rotary Positional Embeddings (RoPE) for improved position encoding. The model features 32 layers, a hidden size of 4096, and 32 attention heads. It employs pre-normalization, Swish-Gated Linear Unit (SwiGLU) activation in feed-forward layers, and various training optimizations. This model was also successfully used in the previous AtomGPT work \cite{choudhary2024atomgpt}.

Now, fine-tuning requires transforming the instructions into a specialized protocol such as Alpaca \cite{taori2023stanford}. The Alpaca instructions consist of Python dictionaries with keys for instruction, input, and output texts. The instruction key was set to ``Below is a description of a material." The XRD patterns were interpolated on a grid of 180 points, with intervals of 0.5 $\degree$ 2$\theta$, using three floating-point precision, and then converted to a string with a newline character as separators. A fixed pattern length allows for uniform token lengths for LLMs, irrespective of different simulation and experimental settings for PXRD data. Note that with decreasing intervals (here 0.5), the number of tokens increases, and hence, the training and inference time will be higher. The input key used was of three types: 1) with no chemical information, 2) with elemental lists only, and 3) with an explicit chemical formula. For the input with no chemical information, the input key was simply ``The XRD is ... Generate atomic structure description with lattice lengths, angles, coordinates, and atom types." Similarly, for the second and third cases, the inputs were ``The chemical elements are ... The XRD is ... Generate atomic structure description with lattice lengths, angles, coordinates, and atom types." and ``The chemical formula is ... The XRD is ... Generate atomic structure description with lattice lengths, angles, coordinates, and atom types," respectively. Finally, the output key was a string of lattice lengths, angles, and chemical coordinates along with three fractional coordinates in XYZ format. Two decimal precision was used for lattice parameters and three decimal precision for coordinates.

As directly fine-tuning such an LLM can be computationally expensive, the PEFT method was used within the Hugging Face ecosystem. Additionally, Transformer Reinforcement Learning (TRL) and RoPE \cite{su2024roformer} were employed to patch the Mistral model with fast LoRA \cite{hu2021lora} weights for reduced memory training. After obtaining the PEFT model, corresponding tokenizer, and Alpaca dataset, supervised fine-tuning tasks were carried out with a batch size of 5, using the AdamW 8-bit optimizer and a cross-entropy loss function for 5 epochs. This loss function measures the difference between the predicted probability distribution over the vocabulary and the true distribution (i.e., the one-hot encoded target words). After the model is trained, it is evaluated on the test set with respect to reconstruction/test performance. To further clarify, after training the model on the training set, while keeping the instruction and input keys in the test set, the trained model is employed to generate outputs. After parsing the outputs to create corresponding crystal structures, the StructureMatcher algorithm \cite{ong2013python} is used to find the best match between two structures, considering all invariances of materials. The root mean square error (RMS) is averaged over all matched materials. Because the interatomic distances can vary significantly for different materials, the RMS is normalized following the work in Ref. \cite{xie2021crystal}. Note that this is just one of the metrics for generative models for atomic structures, and there can be numerous other types of metrics.

In addition to developing GPT models, convolutional neural networks (CNN) and gradient boosting regression tree (GBR) models were developed to predict lattice lengths given XRD patterns, with the same train-test split as for GPT models. For the GBR and CNN models, the XRD signals are used as inputs and the three lattice constants as outputs. For GBR, we used 1000 estimators, a learning rate of 0.01, and a maximum depth of 3 with a mean absolute error loss function. The CNN model used in this study, referred to as CNNRegressor, is designed to perform regression tasks by extracting features from one-dimensional input data. The architecture begins with two 1D convolutional layers: the first layer has 16 filters and the second layer has 32 filters, both with a kernel size of 3 and padding of 1 to preserve the input size. Each convolutional layer is followed by a  Rectified Linear Unit (ReLU) activation function to introduce non-linearity. MaxPooling layers with a kernel size of 2 and stride of 2 are applied to downsample the feature maps, reducing dimensionality and computational load. After these operations, the output is flattened to a shape of 32 × 45, which feeds into a fully connected layer with 64 neurons. The final output layer contains 3 neurons, corresponding to the three target values predicted by the model. This architecture allows the network to efficiently learn relevant features from the input data for accurate regression. The CNN model was trained for 50 epochs with a batch size of 32.

Finally, XRD measurements were also performed for this work to validate the simulated XRD patterns. The crystal structures were characterized using spatially resolved powder X-ray diffraction with a Bruker D8 Discover. We explored Bragg angles ranging from 10 $\degree$ 2$\theta$ to 90 $\degree$ 2$\theta$ using Cu K$\alpha$ radiation (wavelength 1.54184 $\mathring{A}$) at 50 kV, with a step size of 0.02 $\degree$ and a scan rate of 6 $\degree$ per minute.


\begin{figure}[hbt!]
\centering
\includegraphics[width=\linewidth]{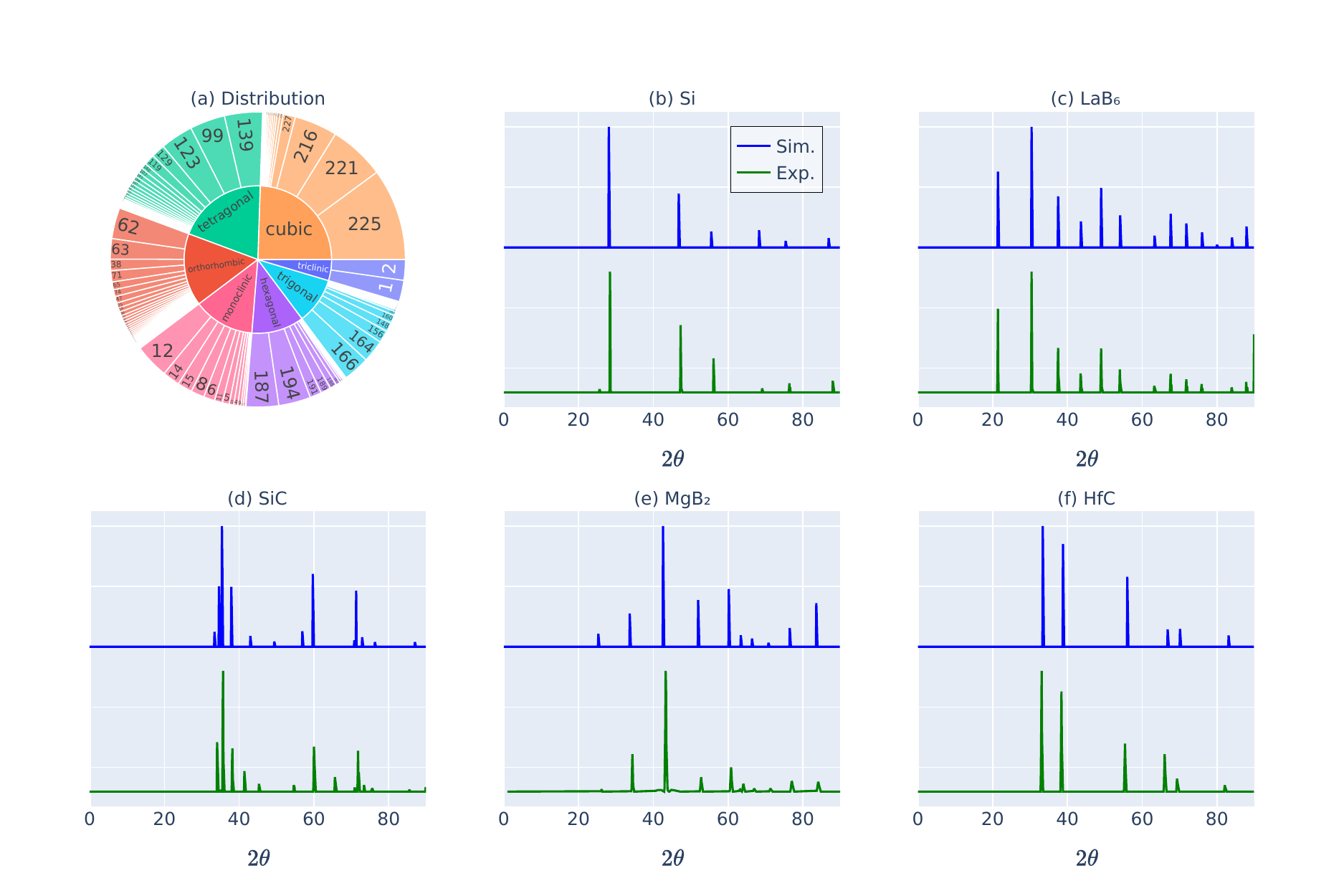}
\caption{Crystal lattice and spacegroup data-distribution in the JARVIS-DFT (JDFT) database and comparison of a few simulated XRD-patterns with experimental measurements. a) Crystal lattice and spacegroup distribution in the JDFT atomic structure database. b) Simulated and experimental PXRD for silicon. The experimental data was taken from RRUFF database with ID R050145 while the simulated data from JDFT ID JVASP-1002, c) Simulated and experimental PXRD for lanthanum boride. The experimental data was obtained as a part of this work while the simulated data from JDFT with ID of 15014, d) Simulated and experimental PXRD for silicon carbide (Moissanite). The experimental data was taken from RRUFF database with ID R061083 while the simulated data from JDFT ID JVASP-107, e) Simulated and experimental PXRD for magnesium boride. The experimental data was obtained as a part of this work while the simulated data from JDFT ID JVASP-1151, f) Simulated and experimental PXRD for hafnium carbide. The experimental data was obtained as a part of this work while the simulated data from JDFT ID JVASP-17957.}
\label{fig:dist}
\end{figure}
In Fig. \ref{fig:dist}, we show the crystal lattice and space group data distribution in the JDFT database and a comparison of several simulated XRD patterns with experimental measurements. In Fig. \ref{fig:dist}a, we observe that most of the crystals are cubic, while the least number belongs to the triclinic lattice out of the seven crystal systems. Similarly, out of 230 space groups, 225, which belong to the cubic lattice system, is prevalent. Such analysis provides a basic understanding of the predictive limits of the models. For instance, if the model is trained with a sufficiently large cubic dataset but not with a triclinic dataset, it might generalize well for cubic systems but not for triclinic ones.

There are various proprietary databases that contain PXRD and atomic structure information. However, in this work, we choose to use the publicly available JARVIS-DFT dataset for proof of concept. Note that although a simulated PXRD database is used here, it can be easily extended to include experimental data in the future. Analyzing the accuracy of the simulated PXRD compared to experimental results is important. In Fig. \ref{fig:dist}b-f, we present a few such comparisons. The experimental data was either obtained from RRUFF database or as part of the experimental component of this work.

The simulated and experimental PXRD for silicon, which is undoubtedly the most important material, especially for the semiconductor industry, is shown in Fig. \ref{fig:dist}b. The experimental data was taken from the RRUFF database with ID R050145, while the simulated data is from JDFT with ID JVASP-1002. All the simulation and experimental data were rescaled between 0 and 1 based on the maximum height available in that pattern for uniform comparison. We can observe close agreement between the simulated (Sim.) and experimental (Exp.) patterns, suggesting high fidelity of the simulated data. We note that the relative peak heights may not be exactly identical for all the peaks, which can be attributed to the collection of crystal planes encountered during PXRD experiments.

Similarly, the simulated and experimental PXRD for lanthanum boride, considered an important reference material for XRD, is shown in Fig. \ref{fig:dist}c. The experimental data was obtained as part of this work, while the simulated data is from JDFT with ID JVASP-15014. Here, we observe excellent agreement in peak positions and peak height values, especially up to  $60^\circ$ 2$\theta$ values, after which peak heights begin to differ. The simulated and experimental PXRD for silicon carbide (Moissanite) is shown in Fig. \ref{fig:dist}d. The experimental data was taken from the RRUFF database with ID R061083, while the simulated data is from JDFT with ID JVASP-107. Here, we see more peaks in the simulation around  $30^\circ$ 2$\theta$, which can also be attributed to the reasons mentioned above regarding crystal planes encountered during experiments. PXRD should measure an aggregate of all present crystal planes that diffract X-ray that fulfill the Braggs criterion. However, in experiments, it is possible to miss some of the plane orientations in the powder sample. Finally, the simulated and experimental PXRDs for magnesium boride and hafnium carbide are shown in Fig. \ref{fig:dist}e-f. In the case of magnesium boride, we are missing a peak around the  $20^\circ$ 2$\theta$ value, as well as peaks after  $60^\circ$ 2$\theta$. We observe excellent agreement in the hafnium carbide case, especially up to  $60^\circ$ 2$\theta$ values, after which the experimental data shows fewer peaks than the simulated data. After generating such PXRD patterns for all the materials in JDFT, we perform LLM training following the details mentioned above, and the resultant models can be used for fast prediction of crystal structures.

As the first evaluation of the model's performance, the lattice constants in the x, y, and z crystallographic directions are compared for crystals in the test set and those generated using the DGPT models. This test set was never exposed to the model during training. The lattice constants from XRD can also be predicted using other ML techniques such as gradient boosting regression tree (GBR), convolutional neural networks (CNN), and various DiffractGPT (DGPT) models, as shown in Table 1. The mean absolute errors (MAE) for predicting a, b, and c lattice constants on the test set for GBR are 1.03 $\mathring{A}$, 0.99 $\mathring{A}$, and 1.27 $\mathring{A}$. Similarly, for CNN models, MAEs of 0.28 $\mathring{A}$, 0.27 $\mathring{A}$, and 0.28 $\mathring{A}$ are observed, which is a significant improvement compared to GBR. Now, the performance of three types of DiffractGPT models—those with chemical information, with element lists, and with explicit formulas—shows the minimum error for the model with explicit formulas, which is intuitively correct. Specifically, the lowest error in lattice constant predictions was observed for the a-lattice parameter at 0.17 $\mathring{A}$. This value is close to the CNN model predictions. Li et al. performed a similar task for predicting lattice constants and found a mean absolute error (MAE) of 0.48 $\mathring{A}$ \cite{li2021mlatticeabc} and an R$^2$ of 0.80. Although the datasets for these two works are different, a MAE of 0.17 $\mathring{A}$ suggests promising results. As larger databases are used for DiffractGPT in the future, the MAE may further decrease. Note that DiffractGPT provides not only lattice constants but also full atomic structure information, such as chemical elements and coordinates. Hence, as a second evaluation, we compare the root mean square distance (RMS-d) between the predicted and target materials in the test set and find that the lowest error is observed for the DGPT model with explicit formulas. The RMS-d of 0.07 $\mathring{A}$ is comparable to the AtomGPT value of 0.08 $\mathring{A}$ for the superconductor design task \cite{choudhary2024atomgpt}.

\begin{figure}[hbt!]
\centering
\includegraphics[width=\linewidth]{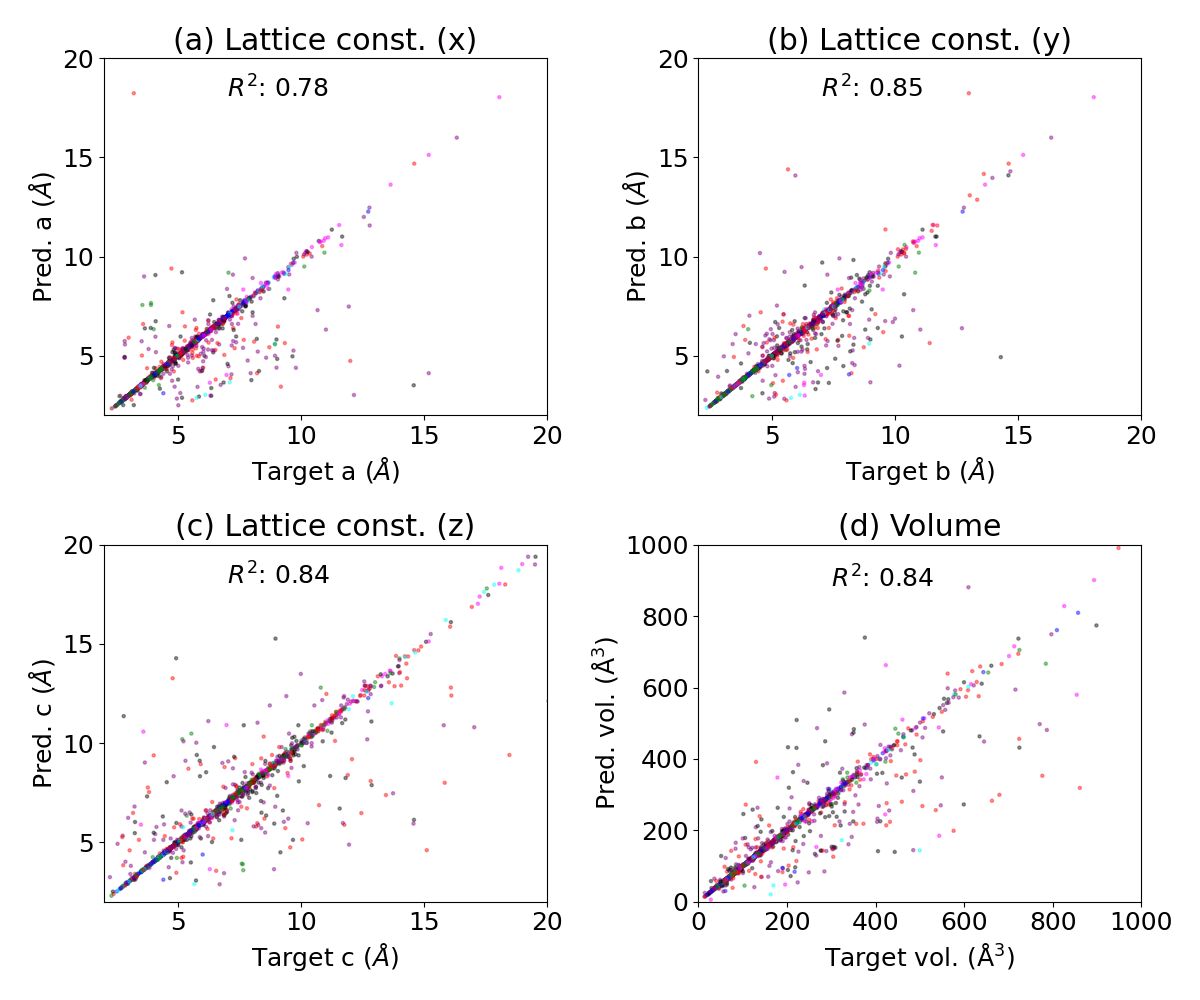}
\caption{Performance of DiffracGPT chemical formula+XRD pattern to atomic structure model for lattice constants in a) x-crystallographic direction, b) y-crystallographic direction, c) z-crystallographic direction, d) volume. The color of the dots in the plot represents different crystal lattice types. The cubic, tetragonal, orthorhombic, hexagonal, trigonal, monoclinic, and triclinic systems are represented by blue, green, red, cyan, magenta, purple, and black colors, respectively. The values that lie on the x = y line represent perfect agreement, while points away from it represent outliers.
}
\label{fig:lattice}
\end{figure}

To illustrate further, we show the predicted lattice constants and volumes for the DiffractGPT chemical formula + XRD pattern model in Fig. \ref{fig:lattice}. The color of the dots in the plot represents different crystal lattice types. The cubic, tetragonal, orthorhombic, hexagonal, trigonal, monoclinic, and triclinic systems are represented by blue, green, red, cyan, magenta, purple, and black colors, respectively. The values that lie on the x = y line represent perfect agreement, while points away from it represent outliers. We barely observe outliers from symmetric lattice systems such as cubic materials. Most of the outliers are from the red and purple dots, representing orthorhombic and monoclinic systems. We find the maximum R$^2$ score of 0.85 (for lattice constant b) and the minimum R$^2$ of 0.78 for lattice constant a.

\begin{table}
\begin{tabular}{cccccc}
\hline

Prop/MAE&GBR&CNN&DGPT-no formula&DGPT-element list&DGPT-formula\\
\hline
a & 1.03 & 0.28 &  0.25 & 0.18& \textbf{0.17}\\
b & 0.99 & 0.27 &0.26 & 0.20&\textbf{0.18} \\
c& 1.27 & 0.28 & 0.38 & 0.28&\textbf{0.27} \\
RMS-d& - & - & 0.23 & 0.21&\textbf{0.07} \\
\hline

\end{tabular}
\caption{Performance measurement in terms of mean absolute error (MAE) for predicting lattice constants ($\mathring{A}$) using gradient boosting regression (GBR), convolutional neural network (CNN), and varieties of DiffractGPT (DGPT) models. We also compare root mean square distance in predicted vs target structures using DGPT models.}

\end{table}

\begin{figure}[hbt!]
\centering
\includegraphics[width=\linewidth]{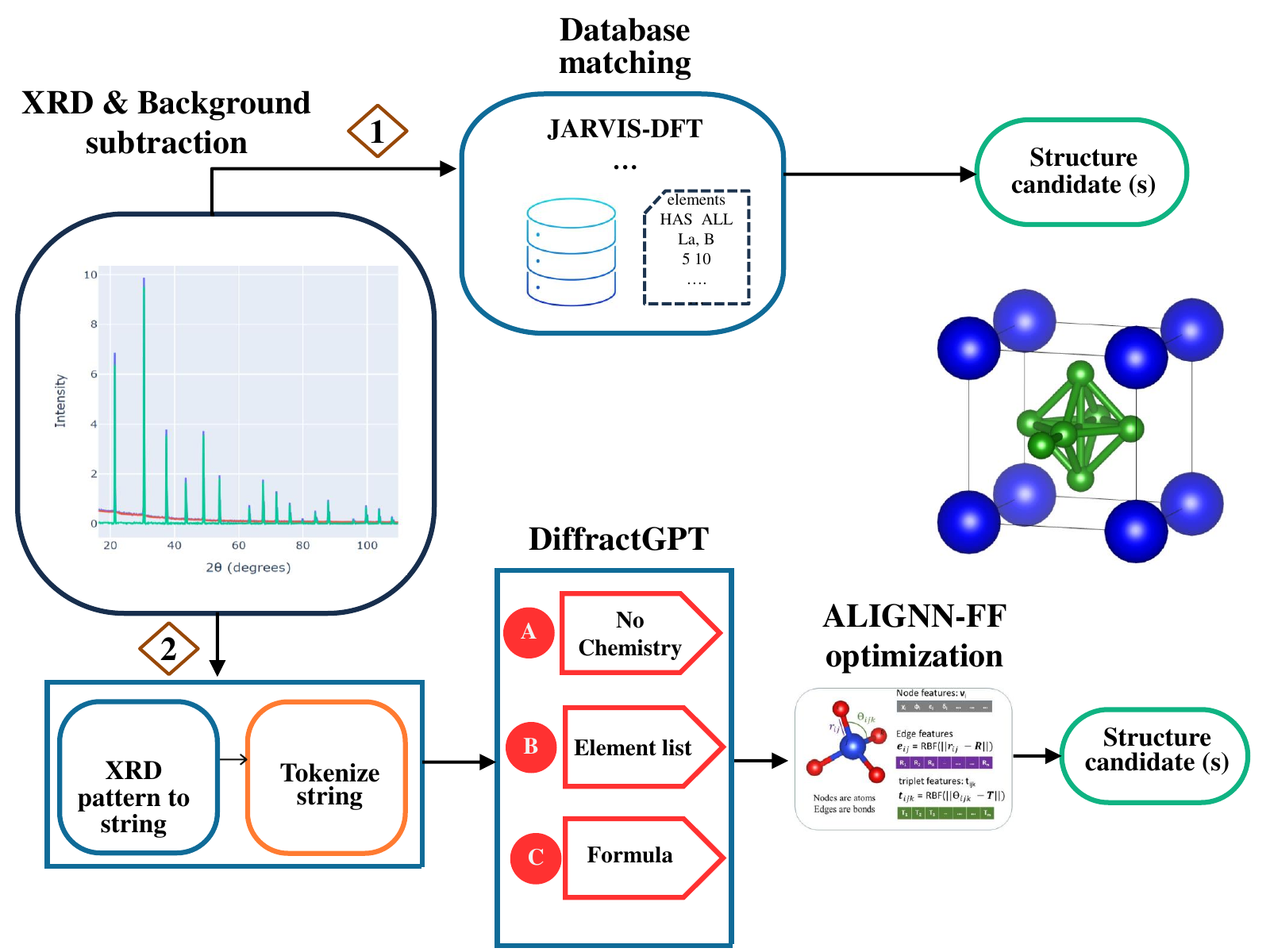}
\caption{Schematic Overview of Crystal Structure Determination from XRD Patterns Using the DiffractGPT Workflow. It begins with the user providing an XRD pattern as input. Utilizing the scripts available in JARVIS-Tools, background subtraction is automatically performed. First, the spectrum can be matched with structures from atomic structure databases, such as those in JDFT or similar databases, based on simulated XRD patterns using cosine similarity or other metrics. Alternatively, there are multiple scenarios where the user might (1) not know the constituent elements at all, (2) have some idea about the involved elements, or (3) explicitly know the chemical formula. Based on the provided information, the XRD pattern can be converted to strings followed by tokenization, after which one or more pre-trained DiffractGPT models can be applied to generate potential crystal structures. Subsequently, further optimization can be performed using a unified GNN force field, such as ALIGNN-FF, to generate additional structure candidates. A tentative application for this workflow is available at the website https://jarvis.nist.gov/jxrd.}
\label{fig:schematic}
\end{figure}

\begin{figure}[hbt!]
\centering
\includegraphics[width=\linewidth]
{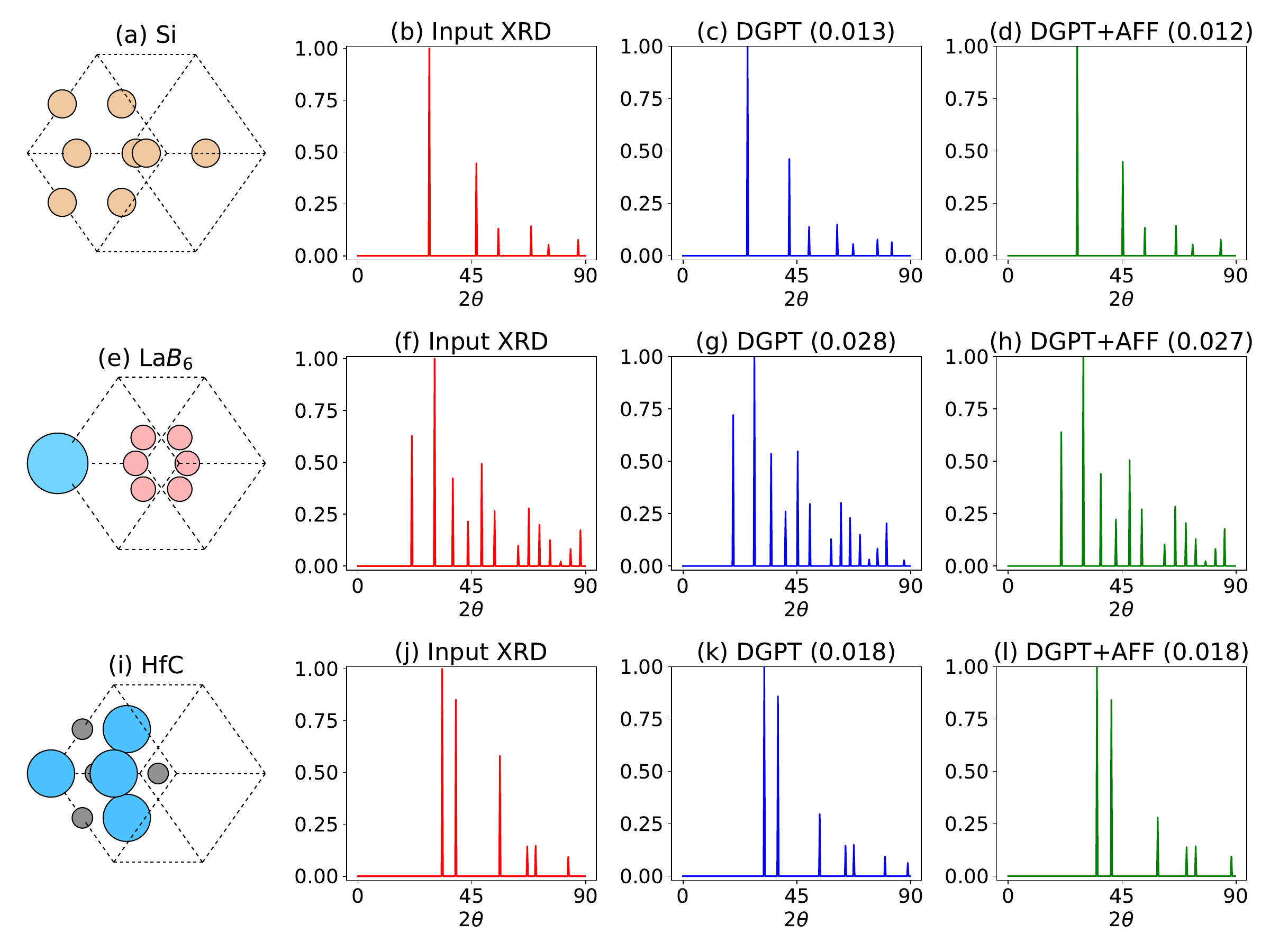}  

\caption{\textcolor{black}{Evaluating the performance of DiffractGPT (DGPT)-formula model with and without ALIGNN-FF (AFF) optimization for a few example materials. The input chemical formula and XRD pattern are fed into the DGPT model to generate the atomic structure. The theoretical XRD pattern of the generated structure is shown as DGPT, along with the mean absolute error (MAE) of the XRD pattern in comparison with the input XRD. The DGPT structure is further optimized with AFF, and the XRD of the optimized structure, along with its MAE, is shown. (a) Silicon atomic structure, (b) input XRD pattern for Si, (c) XRD pattern of the DGPT-generated structure given the chemical formula and XRD, (d) XRD pattern for the AFF-optimized DGPT structure. Similar results for LaB6 (e-h) and HfC (i-l) are shown.}}
\label{fig:comp}
\end{figure}

\begin{figure}[hbt!]
\centering
\includegraphics[width=\linewidth]{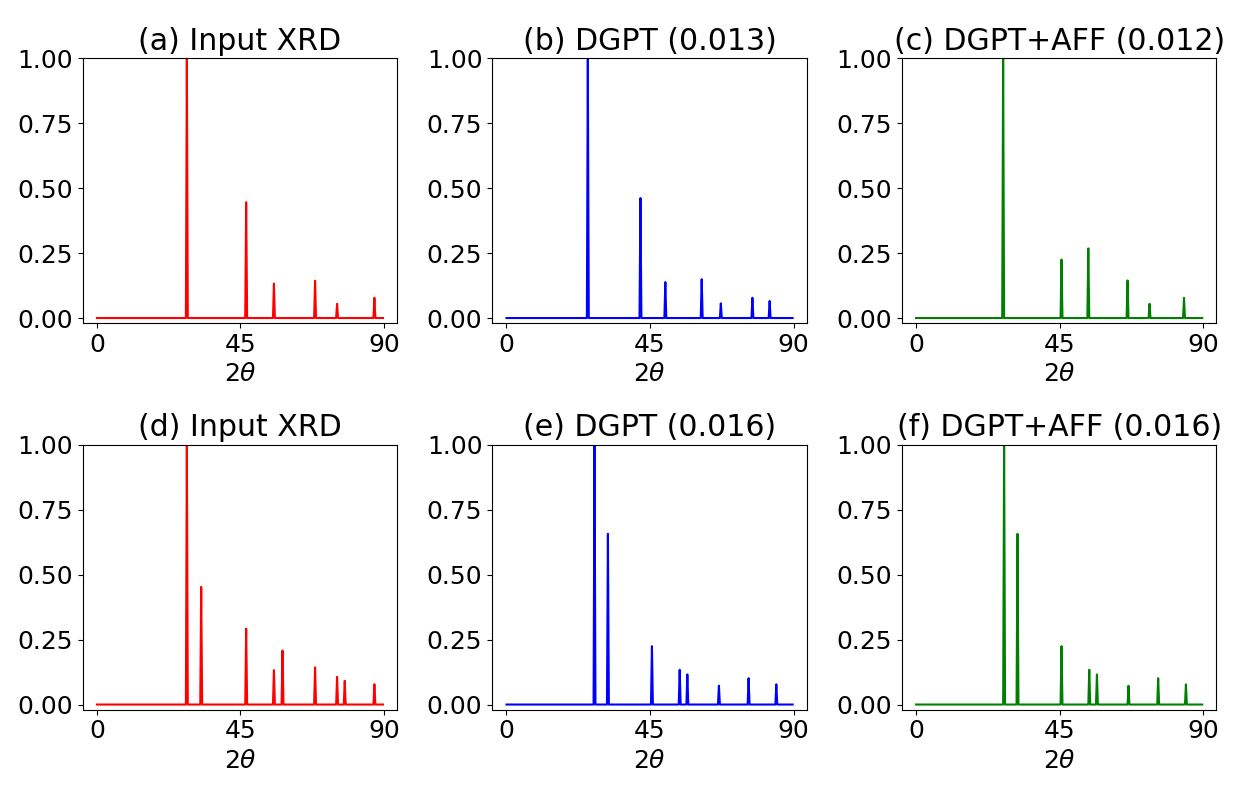}

\caption{XRD patterns for both perfect and defective two-atom Silicon (JVASP-1002) structure, with and without the displacement of an atom from its equilibrium position, are shown. The x-coordinate of the first atom is translated by 0 (panels a-c) and 0.2 (panels d-f), with the 0 translation representing the perfect crystal. After generating the crystals, we predict their simulated patterns. We then use these patterns, along with the chemical formula Si, to generate the DGPT-based atomic structure and its corresponding diffraction pattern. Furthermore, the DGPT-generated structure is optimized using ALIGNN-FF, and the corresponding XRD patterns are also presented.}
\label{fig:transl}
\end{figure}

Now, we present an overview of the usability of the DiffractGPT framework in Fig. \ref{fig:schematic}. DiffractGPT can be used to predict the complete crystal structure given a PXRD pattern. A user provides a PXRD pattern as input. These patterns contain background noise, which can be automatically detected and subtracted using scripts available in JARVIS-Tools. As a first option, the spectrum can be matched with structures from atomic structure databases, such as those in JDFT or similar databases, based on simulated XRD patterns using cosine similarity or other metrics. A web application for this option is available at the JARVIS-XRD website (\url{https://jarvis.nist.gov/jxrd}). This process can predict the top candidates for the input XRD pattern. However, if the XRD patterns are complex or if the material does not exist in the current databases, the second option can be employed as follows. There are multiple scenarios: the user might (1) not know the constituent chemical elements at all, (2) have some idea about the involved elements, or (3) explicitly know the chemical formula. We have independent DiffractGPT models for all these scenarios. Based on the provided information, we can convert the XRD pattern to strings followed by tokenization, after which one or more pre-trained DiffractGPT models can be applied to generate potential crystal structures. Note that transformer architectures allow for fast sampling, which can also be used to generate multiple options for the crystal structure if necessary.

As an optional subsequent step, further optimization of the generated structures can be performed using a unified graph neural network (GNN) force field (FF), such as the atomistic line graph neural network (ALIGNN)-FF \cite{choudhary2023unified}, to generate additional structure candidates. It was developed for fast crystal structure optimization and to handle chemically and structurally diverse crystalline systems, with the entirety of the JARVIS-DFT dataset used for training. This dataset contains 4 million energy-force entries for 89 elements of the periodic table, of which 307,113 entries were utilized for training \cite{choudhary2023unified}. ALIGNN-FF is seamlessly integrated into the DiffractGPT framework.

\textcolor{black}{In Fig. \ref{fig:comp}, we evaluate the performance of the DiffractGPT (DGPT)-formula model with and without ALIGNN-FF (AFF) optimization for a few selected materials. In these examples, the input chemical formula and X-ray diffraction (XRD) pattern are fed into the DGPT model to generate an initial atomic structure. The theoretical XRD pattern of the generated structure is shown, along with the mean absolute error (MAE) when compared to the original input XRD pattern. To further demonstrate the impact of optimization, we apply the ALIGNN-FF (AFF) force field to relax the DGPT-generated structure, and the resulting XRD pattern for the optimized structure is shown along with its corresponding MAE. We observe some of the limitations of the model. For example, in Fig. \ref{fig:comp}a, there are 6 peaks while the DGPT model generates model with 7 peaks for Silicon as shown in Fig. \ref{fig:comp}b. After applying the ALIGNN-FF optimization, we observe that the number of peaks is corrected to 6, as expected as shown in Fig. \ref{fig:comp}c. A similar trend is observed for LaB$_6$, where the input XRD pattern has 13 peaks (Fig. \ref{fig:comp}f), but the DGPT model initially predicts 14 peaks (Fig. \ref{fig:comp}g). This discrepancy is also corrected with ALIGNN-FF optimization. On the other hand, for the HfC case shown in Fig. \ref{fig:comp}j, the predicted XRD pattern consistently matches the correct number of peaks, suggesting that ALIGNN-FF optimization may not be necessary in this case. We further quantify these observations with mean absolute error (MAE) values, comparing the target and predicted XRD patterns. The structure with the lower MAE can be considered the better candidate structure for the XRD pattern. Moreover, while for the above analysis simulated XRD patterns were used as inputs, the same for experimental patterns is shown in Fig. S1.  The experimental XRD pattern is scaled between 0 and 1 and peaks less than 0.04 as a threshold value are removed to align with the training based simulated data. Interestingly, we observe excellent agreement for Si and LaB$_6$ case, but for HfC case we observe noticeable difference.} 

\textcolor{black}{While the above analysis provides insights into the performance of the model in different scenarios, obtaining deeper physical insights into why these discrepancies occur is a more complex task. Due to the nature of deep learning models with billions of parameters, they tend to be less explainable, making it difficult to extract detailed physical explanations. However, we plan to explore such investigations in future work to better understand these behaviors.}

\textcolor{black}{Furthermore, there could be different types of real world diffraction patterns including defects. An example of silicon structure with and without defects (translated atom) is shown in Fig. \ref{fig:transl}. After constructing a perfect silicon structure with two atoms in the primitive cell, The x-coordinate of the first atom is translated by 0 (panels a-c) and 0.2 (panels d-f), with the 0 translation representing the perfect crystal. After generating the crystals, we predict their simulated patterns. We then use these patterns, along with the chemical formula Si, to generate the DGPT-based atomic structure and its corresponding diffraction pattern. Furthermore, the DGPT-generated structure is optimized using ALIGNN-FF, and the corresponding XRD patterns are also presented. We observe that for the defective structure, the peaks show reasonable agreement for peaks before $45^\circ$ 2$\theta$, but after that, it begins to differ compared to input XRD pattern. This can be attributed to the fact that the current work has primarily focused on perfect materials with no defective structures explicitly included during training. However, it could be extended to defective materials in the future. Detecting defects, such as vacancies, dislocations or other imperfections, in materials through X-ray diffraction (XRD) is a challenging task. While XRD is commonly used to study crystalline materials, the presence of defects introduces complexities in the diffraction patterns. Previous studies, such as those utilizing convolutional neural networks \cite{lim2021convolutional,boulle2023convolutional,judge2023defect} and Long Short-Term Memory (LSTM) networks \cite{motamedi2024lstm} for identifying vacancies, strain in semiconductors, have made progress in this area. Our model, trained on diffraction patterns from ideal structures, can be extended to defective systems by incorporating additional training data from materials with known defects. With such data, the model should be able to generalize and capture the diffraction features associated with defects and dislocations. }


In conclusion, this study introduces an efficient approach for determining crystal structures from powder X-ray diffraction patterns. It goes beyond existing generative AI applications focused on scalar properties by facilitating structure generation and demonstrating the potential of using spectral data, such as XRD. The DiffractGPT model is capable of predicting material properties with high accuracy, particularly when the chemical elements of the materials are known. Notably, DiffractGPT outperforms conventional machine learning models, such as gradient boosting and convolution neural network, in predicting lattice constants while also providing the option to generate complete crystal structures. Additionally, the training process for DiffractGPT is straightforward, fast and relatively easy to learn, thereby bridging the gap between the computational, data science, and experimental communities. As a complementary tool, we offer a framework that matches experimental XRD patterns with existing databases, incorporating automated background subtraction. This work represents a significant advancement in the automation of crystal structure determination and provides a robust tool for data-driven materials design, paving the way for enhanced research and development in materials science.

\section*{Supporting Information}
Additional examples of evaluating the performance of DiffractGPT-formula model with experimental
XRD patterns as inputs.

\section*{Acknowledgements}

K.C. thanks computational resources from the National Institute of Standards and Technology (NIST). K.C. thanks Maureen E. Williams, Adam J. Biacchi and Adam A. Creuziger at NIST for helpful discussion. This work was performed with funding from the CHIPS Metrology Program, part of CHIPS for America, National Institute of Standards and Technology, U.S. Department of Commerce. 
Certain commercial equipment, instruments, software, or materials are identified in this paper in order to specify the experimental procedure adequately. Such identifications are not intended to imply recommendation or endorsement by NIST, nor it is intended to imply that the materials or equipment identified are necessarily the best available for the purpose.

\bibliography{achemso-demo}

\end{document}